\RequirePackage{lineno}
\documentclass[aps,prl,twocolumn,showpacs,superscriptaddress,groupedaddress]{revtex4}  % for review and submission
\usepackage{graphicx}  % needed for figures
\usepackage{dcolumn}   % needed for some tables
\usepackage{bm}        % for math
\usepackage{amssymb}   % for math
\usepackage{comment}   % for commenting out entire sections

\def\MET{{\mbox{$p\kern-0.42em\raise0.15ex\hbox{/}_{T}$}}}
\def\met{{\mbox{$p\kern-0.42em\raise0.15ex\hbox{/}_{T}$}}}
\def\mex{{\mbox{$p\kern-0.42em\raise0.15ex\hbox{/}_{x}$}}}
\def\mey{{\mbox{$p\kern-0.42em\raise0.15ex\hbox{/}_{y}$}}}

\def\Dzero{D0~}

\def\ifb{fb$^{-1}~$}

\def\tt{$t\bar{t}~$}

\def\to{\rightarrow}

\def\lmet{$WH\rightarrow \ell\kern-0.45em\raise0.19ex\hbox{/} \nu b\bar{b}~$}

\def\sherpa{\textsc{sherpa~}}

\def\Blackhat{\textsc{blackhat+sherpa~}}

\begin{document}
%\setpagewiselinenumbers
%\modulolinenumbers[5]
%\linenumbers

% the following line is for submission, including submission to the arXiv!!
\hspace{5.2in} \mbox{FERMILAB-PUB-11-246-E-PPD}

\title{Measurements of inclusive W+jets production rates as a function of jet transverse momentum in {\boldmath $p\bar p$} collisions at {\boldmath $\sqrt{s}=1.96$ TeV}}
\affiliation{Universidad de Buenos Aires, Buenos Aires, Argentina}
\affiliation{LAFEX, Centro Brasileiro de Pesquisas F{\'\i}sicas, Rio de Janeiro, Brazil}
\affiliation{Universidade do Estado do Rio de Janeiro, Rio de Janeiro, Brazil}
\affiliation{Universidade Federal do ABC, Santo Andr\'e, Brazil}
\affiliation{Instituto de F\'{\i}sica Te\'orica, Universidade Estadual Paulista, S\~ao Paulo, Brazil}
\affiliation{Simon Fraser University, Vancouver, British Columbia, and York University, Toronto, Ontario, Canada}
\affiliation{University of Science and Technology of China, Hefei, People's Republic of China}
\affiliation{Universidad de los Andes, Bogot\'{a}, Colombia}
\affiliation{Charles University, Faculty of Mathematics and Physics, Center for Particle Physics, Prague, Czech Republic}
\affiliation{Czech Technical University in Prague, Prague, Czech Republic}
\affiliation{Center for Particle Physics, Institute of Physics, Academy of Sciences of the Czech Republic, Prague, Czech Republic}
\affiliation{Universidad San Francisco de Quito, Quito, Ecuador}
\affiliation{LPC, Universit\'e Blaise Pascal, CNRS/IN2P3, Clermont, France}
\affiliation{LPSC, Universit\'e Joseph Fourier Grenoble 1, CNRS/IN2P3, Institut National Polytechnique de Grenoble, Grenoble, France}
\affiliation{CPPM, Aix-Marseille Universit\'e, CNRS/IN2P3, Marseille, France}
\affiliation{LAL, Universit\'e Paris-Sud, CNRS/IN2P3, Orsay, France}
\affiliation{LPNHE, Universit\'es Paris VI and VII, CNRS/IN2P3, Paris, France}
\affiliation{CEA, Irfu, SPP, Saclay, France}
\affiliation{IPHC, Universit\'e de Strasbourg, CNRS/IN2P3, Strasbourg, France}
\affiliation{IPNL, Universit\'e Lyon 1, CNRS/IN2P3, Villeurbanne, France and Universit\'e de Lyon, Lyon, France}
\affiliation{III. Physikalisches Institut A, RWTH Aachen University, Aachen, Germany}
\affiliation{Physikalisches Institut, Universit{\"a}t Freiburg, Freiburg, Germany}
\affiliation{II. Physikalisches Institut, Georg-August-Universit{\"a}t G\"ottingen, G\"ottingen, Germany}
\affiliation{Institut f{\"u}r Physik, Universit{\"a}t Mainz, Mainz, Germany}
\affiliation{Ludwig-Maximilians-Universit{\"a}t M{\"u}nchen, M{\"u}nchen, Germany}
\affiliation{Fachbereich Physik, Bergische Universit{\"a}t Wuppertal, Wuppertal, Germany}
\affiliation{Panjab University, Chandigarh, India}
\affiliation{Delhi University, Delhi, India}
\affiliation{Tata Institute of Fundamental Research, Mumbai, India}
\affiliation{University College Dublin, Dublin, Ireland}
\affiliation{Korea Detector Laboratory, Korea University, Seoul, Korea}
\affiliation{CINVESTAV, Mexico City, Mexico}
\affiliation{Nikhef, Science Park, Amsterdam, the Netherlands}
\affiliation{Radboud University Nijmegen, Nijmegen, the Netherlands and Nikhef, Science Park, Amsterdam, the Netherlands}
\affiliation{Joint Institute for Nuclear Research, Dubna, Russia}
\affiliation{Institute for Theoretical and Experimental Physics, Moscow, Russia}
\affiliation{Moscow State University, Moscow, Russia}
\affiliation{Institute for High Energy Physics, Protvino, Russia}
\affiliation{Petersburg Nuclear Physics Institute, St. Petersburg, Russia}
\affiliation{Instituci\'{o} Catalana de Recerca i Estudis Avan\c{c}ats (ICREA) and Institut de F\'{i}sica d'Altes Energies (IFAE), Barcelona, Spain}
\affiliation{Stockholm University, Stockholm and Uppsala University, Uppsala, Sweden}
\affiliation{Lancaster University, Lancaster LA1 4YB, United Kingdom}
\affiliation{Imperial College London, London SW7 2AZ, United Kingdom}
\affiliation{The University of Manchester, Manchester M13 9PL, United Kingdom}
\affiliation{University of Arizona, Tucson, Arizona 85721, USA}
\affiliation{University of California Riverside, Riverside, California 92521, USA}
\affiliation{Florida State University, Tallahassee, Florida 32306, USA}
\affiliation{Fermi National Accelerator Laboratory, Batavia, Illinois 60510, USA}
\affiliation{University of Illinois at Chicago, Chicago, Illinois 60607, USA}
\affiliation{Northern Illinois University, DeKalb, Illinois 60115, USA}
\affiliation{Northwestern University, Evanston, Illinois 60208, USA}
\affiliation{Indiana University, Bloomington, Indiana 47405, USA}
\affiliation{Purdue University Calumet, Hammond, Indiana 46323, USA}
\affiliation{University of Notre Dame, Notre Dame, Indiana 46556, USA}
\affiliation{Iowa State University, Ames, Iowa 50011, USA}
\affiliation{University of Kansas, Lawrence, Kansas 66045, USA}
\affiliation{Kansas State University, Manhattan, Kansas 66506, USA}
\affiliation{Louisiana Tech University, Ruston, Louisiana 71272, USA}
\affiliation{Boston University, Boston, Massachusetts 02215, USA}
\affiliation{Northeastern University, Boston, Massachusetts 02115, USA}
\affiliation{University of Michigan, Ann Arbor, Michigan 48109, USA}
\affiliation{Michigan State University, East Lansing, Michigan 48824, USA}
\affiliation{University of Mississippi, University, Mississippi 38677, USA}
\affiliation{University of Nebraska, Lincoln, Nebraska 68588, USA}
\affiliation{Rutgers University, Piscataway, New Jersey 08855, USA}
\affiliation{Princeton University, Princeton, New Jersey 08544, USA}
\affiliation{State University of New York, Buffalo, New York 14260, USA}
\affiliation{Columbia University, New York, New York 10027, USA}
\affiliation{University of Rochester, Rochester, New York 14627, USA}
\affiliation{State University of New York, Stony Brook, New York 11794, USA}
\affiliation{Brookhaven National Laboratory, Upton, New York 11973, USA}
\affiliation{Langston University, Langston, Oklahoma 73050, USA}
\affiliation{University of Oklahoma, Norman, Oklahoma 73019, USA}
\affiliation{Oklahoma State University, Stillwater, Oklahoma 74078, USA}
\affiliation{Brown University, Providence, Rhode Island 02912, USA}
\affiliation{University of Texas, Arlington, Texas 76019, USA}
\affiliation{Southern Methodist University, Dallas, Texas 75275, USA}
\affiliation{Rice University, Houston, Texas 77005, USA}
\affiliation{University of Virginia, Charlottesville, Virginia 22901, USA}
\affiliation{University of Washington, Seattle, Washington 98195, USA}
\author{V.M.~Abazov} \affiliation{Joint Institute for Nuclear Research, Dubna, Russia}
\author{B.~Abbott} \affiliation{University of Oklahoma, Norman, Oklahoma 73019, USA}
\author{B.S.~Acharya} \affiliation{Tata Institute of Fundamental Research, Mumbai, India}
\author{M.~Adams} \affiliation{University of Illinois at Chicago, Chicago, Illinois 60607, USA}
\author{T.~Adams} \affiliation{Florida State University, Tallahassee, Florida 32306, USA}
\author{G.D.~Alexeev} \affiliation{Joint Institute for Nuclear Research, Dubna, Russia}
\author{G.~Alkhazov} \affiliation{Petersburg Nuclear Physics Institute, St. Petersburg, Russia}
\author{A.~Alton$^{a}$} \affiliation{University of Michigan, Ann Arbor, Michigan 48109, USA}
\author{G.~Alverson} \affiliation{Northeastern University, Boston, Massachusetts 02115, USA}
\author{G.A.~Alves} \affiliation{LAFEX, Centro Brasileiro de Pesquisas F{\'\i}sicas, Rio de Janeiro, Brazil}
% ...
%
% visitor_addresses.tex                        2 June 2011
%  available symbols are:
%  $\ast, \dag, \ddag, \S, \P, $\|$, $\ast\ast$, \dag\dag, \ddag\ddag ,\#
%
\collaboration{The D0 Collaboration\footnote{with visitors from
%{alton}
$^{a}$Augustana College, Sioux Falls, SD, USA,
%{burdin}
$^{b}$The University of Liverpool, Liverpool, UK,
%{haas,partridge}
$^{c}$SLAC, Menlo Park, CA, USA,
%{hesketh}
$^{d}$University College London, London, UK,
%{luna-garcia}
$^{e}$Centro de Investigacion en Computacion - IPN, Mexico City, Mexico,
%{podesta-lerma}
$^{f}$ECFM, Universidad Autonoma de Sinaloa, Culiac\'an, Mexico,
and 
%{weber}
$^{g}$Universit{\"a}t Bern, Bern, Switzerland.
%{garcia-guerra}
%$^{?}$UPIITA-IPN, Mexico City, Mexico,
%{hooper}
%$^{?}$Visitor from Bradley University, Peoria, IL, USA.
%{kozminski}
%$^{?}$}Visitor from Lewis University, Romeoville, IL, USA.
%{deceased}
$^{\ddag}$Deceased.
}} \noaffiliation
\vskip 0.25cm

\date{August 10, 2011}

\begin{abstract}
This Letter describes measurements of inclusive $W (\rightarrow e \nu) + n$ jet cross sections ($n=$1--4), 
presented as total inclusive cross sections and differentially in the $n^{\text{th}}$ jet transverse momentum. 
The measurements are made using data corresponding to an integrated luminosity of 4.2 \ifb collected by the \Dzero detector at the Fermilab Tevatron Collider, 
and achieve considerably smaller uncertainties on $W$+jets production cross sections than previous measurements.
The measurements are compared to next-to-leading order perturbative QCD (pQCD) calculations in the $n=$1--3 jet multiplicity bins and to leading order pQCD calculations in the 4-jet bin.
The measurements are generally in agreement with pQCD calculations, 
although certain regions of phase space are identified where these predictions could better match the data.
\end{abstract}

\maketitle

Measurements of vector boson plus jet production are fundamental tests of perturbative quantum chromodynamics (pQCD), the theory describing the strong interaction.
In addition to providing a test of pQCD at high momentum scales, $W$+jets production can be the dominant background in measurements of single top quark and $t\bar{t}$ production as well as in searches for the standard model Higgs boson and for physics beyond the standard model.  
Theoretical uncertainties on the production rates and kinematics introduce limitations in
our ability to identify new physics signals.
Therefore, it is crucial to make precision measurements of $W$+jets production at the Fermilab Tevatron Collider and the CERN Large Hadron Collider in order to constrain these backgrounds.
We present new measurements of $W$+jets cross sections with a data sample more than ten times larger than that used in previous measurements~\cite{Aaltonen:2007ip}, allowing the first detailed study of $W+4$ jet production.  The previous measurements have been used extensively in testing and tuning theoretical models of $W$ boson production~\cite{Berger:2009ep,Ellis:2009bu,Gleisberg:2008ta}.

The strategy employed for this measurement is based on those used in the \Dzero 
$Z$+jet cross section~\cite{Abazov:2008ez} and $Z$ boson $p_T$~\cite{Abazov:2010kn} publications.  
We select a high purity sample of $W$+jets events
and the results are corrected to the ``particle level," which includes energy from stable particles, the underlying event, muons, and neutrinos, as defined in Ref.~\cite{particle}.  
This procedure corrects a measured observable back to the particle level observable, correcting for the effect of finite experimental resolution, detector response, acceptance, and efficiencies. 

These measurements use a sample of $W(\to e \nu) + n$ jet candidate events 
corresponding to an integrated luminosity of 4.2 \ifb collected 
with the D0 detector in Run II of the Fermilab Tevatron Collider.  
The D0 detector consists of a central tracking system, comprising a silicon microstrip tracker and a fiber tracker, both within an approximately 2~T axial magnetic field.  These components are used primarily to identify the location of the $p\bar{p}$ interaction vertex and the electron produced in the decay of the $W$ boson candidate.  Outside of the tracking system, a liquid-argon and uranium calorimeter is divided into a central section and two end sections that are used to identify electromagnetic and hadronic showers.
A detailed description of the \Dzero detector can be found in Ref.~\cite{d0det}.

The data were collected using a suite of electron and electron+jet triggers. The lowest electron transverse energy threshold in the electron suite is 22 GeV, and the electron threshold 
for the e+jets triggers is 15~GeV.  The combination of the triggers used provides $>97$\% trigger efficiency for electrons with transverse energy above 26~GeV.  
The efficiency in the turn on region below this energy threshold is evaluated using unbiased data samples and a corresponding scale factor is then applied to the MC simulation.

The events were then processed through the \Dzero reconstruction program which identifies jet and $W$ boson candidates.  
Jets are identified with the \Dzero midpoint cone algorithm~\cite{d0jets}, which uses a cone of radius ${\cal R} = 0.5$ (distance in $\eta-\phi$ space~\cite{definitions}) 
to cluster calorimeter cells.  
The electromagnetic fraction of the jet energy is required to be below $0.95$ to reject electrons and above $0.05$ to suppress jets dominated by noise.
% from the hadronic part of the calorimeter.  
Jets with a large fraction of their energy deposited in the coarse hadronic layers of the calorimeter are also rejected due to noise typical in those layers. 
To minimize background from jet candidates arising from noise in the precision readout of the calorimeter, confirmation from the readout system of the first level 
trigger is required for reconstructed jets.  Jets matched to loose electrons with $p_T > 20$~GeV and $\Delta R(e,\textrm{jet}) < 0.5$ are also rejected.  
Jets are corrected for calorimeter response, instrumental and out-of-cone showering effects, and additional energy deposits in the calorimeter that arise from detector noise and pile-up from multiple interactions and different beam crossings. 
These jet energy scale corrections~\cite{JES} are determined using transverse momentum imbalance in $\gamma$~+~jet
events, where the electromagnetic calorimeter response is calibrated using $Z/\gamma^* \rightarrow e^+e^-$ events.
Jets are required to have at least two tracks that point to their associated $p\bar{p}$ vertex.
Energies of jets containing muons are corrected with the measured muon momentum after accounting for the typical energy deposited by a minimum ionizing particle.
Jets are ordered in decreasing transverse momentum and we call the jet with the highest transverse momentum ``leading."
Electrons are identified as clusters of calorimeter cells in which 95\% of the energy in the shower is deposited in the electromagnetic (EM) section. 
The electron candidates must be isolated from other calorimeter energy deposits, have spatial distributions consistent with those expected for electron showers, 
and the event must contain a reconstructed track matched to the EM shower that is isolated from other tracks.
Isolation from energy deposited by hadrons is imposed by requiring $(E_{\rm tot} - E_{\rm em})/E_{\rm em} < 0.15$, where $E_{\rm tot}$ ($E_{\rm em}$)  
is the total (electromagnetic) energy in a cone of radius $\mathcal{R} = 0.4$ ($\mathcal{R} = 0.2$).
Events with a second isolated electron (with $p_T>15$~GeV) are removed to suppress the background due to $Z$ boson and Drell-Yan production.  
The missing transverse energy in the event is calculated as the vector sum of the calorimeter cell energies
and is corrected for the presence of any muons. 
Because the longitudinal component of the momentum of the neutrino is not measured, the measured properties of the $W$ boson candidates are limited to their transverse energy, $E_T^W$, and transverse mass, defined as 
\begin{equation}
M_T^W = \sqrt{(\met+p_T^e)^2 - (\mex + p_x^e)^2 - (\mey + p_y^e)^2}
\label{Eqn1}
\end{equation}
where \met\ is the magnitude of the missing transverse energy vector, $p_T^e$ is the transverse momentum of the electron, and $p_x^e$ and $p_y^e$ ($\mex$ and $\mey$) 
are the magnitude of the $x$ and $y$ components of the electron's momentum (missing transverse energy) respectively.

The following requirements are used in order to suppress background while maintaining high efficiency for events in which a $W$ boson is produced: $p_T^e \ge 15$ GeV and electron 
pseudorapidity $|\eta^e| < $ 1.1, \met $>$ 20 GeV, $M_T^W \ge$ 40 GeV, jet transverse momentum $p_T^{\text{jet}} \ge$ 20 GeV and rapidity $|y^{\text{jet}}| < $ 3.2, 
$\Delta {\cal R}$ = $\sqrt{(\Delta \phi)^2 + (\Delta \eta)^2}$ between the electron and the nearest jet $>0.5$, and the $z$ component of the $p\bar{p}$ interaction vertex 
is restricted to $|z_{\text{vtx}}| < 60$ cm~\cite{definitions}.
Events must have a reconstructed $p \bar{p}$ interaction vertex, containing at least three associated tracks.
This $p\bar{p}$ interaction vertex is required to be less than 1 cm away in the coordinate along the beam line from the extrapolated electron track.

After these requirements, $W$(+jets) events dominate the data sample but there are backgrounds from $Z$+jets, $W(\to\tau\nu \to e\nu\nu)$+jets, $t\bar{t}$, diboson, 
single top quarks, and multijet events.  
We simulate the $W/Z$+jets and \tt processes with \textsc{alpgen}~\cite{Mangano:2002ea} interfaced with \textsc{pythia}~\cite{pythia} for the simulation of initial and final state radiation and for parton hadronization.
The \textsc{pythia} generator is used to simulate diboson production, while production of single top quarks is simulated with the \textsc{comphep}~\cite{Boos:2004kh} generator interfaced with \textsc{pythia}.
The cross sections for $W/Z$+jet production are taken from \textsc{alpgen}, corrected with a constant multiplicative factor to match the inclusive $W/Z$+jet cross sections calculated at NLO~\cite{MCFM}. 
Additional corrections are applied to events containing $W/Z$ bosons plus heavy flavor jets, to match the predictions of NLO QCD calculations.
Events from randomly chosen beam crossings, with the same instantaneous luminosity profile as the data, are overlaid on the simulated events to reproduce the effect of multiple $p \bar{p}$ interactions and detector noise.
All simulated samples are passed through the D0 detector simulation and then reconstructed in the same way as the data.
The estimated fraction of the data sample that is due to processes other than $W$+jets ranges within 2--40\%. Leptonic background from $W(\to\tau\nu \to e\nu\nu)$+jets processes represents 
approximately 5--8\% of all reconstructed $W$+jets events, and the fraction of background due to top quark production ranges within 0 to 7\% (16\%) in the one (two) jet multiplicity bin, 5--40\% in the three jet bin and 20--60\% in the four jet bin (with the extremes only being reached at the highest jet $p_T$ bins in all cases).

In multijet events, there is a small but non-negligible chance that a jet may be misidentified as an electron and then the event may pass all selection criteria.  
As the multijet cross section is large, the contribution from such instances of fake-electron events to the measured distributions must be taken into account.  To determine the number and kinematic distributions of such events, we use the data-driven method described in Ref.~\cite{Abazov:2007kg} because the estimation of this background from Monte Carlo simulations is not reliable.   This approach uses data in a control region that has no overlap with the signal selection to determine the differential distribution and overall normalization of the multijet distributions.

The total background contribution is subtracted from the data in each bin of the $p_T^{\text{jet}}$ distribution.
After background subtraction, the data are corrected for detector resolution effects using a regularized inversion of the resolution matrix as implemented in the program \textsc{guru}~\cite{Hocker:1995kb}, with ensemble testing used to derive statistical uncertainties and unfolding biases. 
This method is described in detail in Ref.~\cite{Abazov:2010kn}.
We have chosen the matrix unfolding approach over the traditional bin-by-bin correction method because of non-negligible bin migration effects in the $p_T^{\text{jet}}$ variable and because the matrix unfolding method provides improved estimation of the uncertainties of the measurement.

To evaluate statistical uncertainties on the unfolded distributions, as well as systematic biases and uncertainties, we build ensembles using \textsc{alpgen+pythia} signal events 
that have the same statistical fluctuations as the data sample.  
The ensembles are reweighted to accurately describe the kinematics of the unfolded jet $p_T$.  
Five hundred ensembles are created and unfolded in the same manner as the data and are in-turn compared to their corresponding generator-level distributions.  
The residual differences between the generator-level and unfolded measurement in each bin, for each ensemble, are determined and fitted with a Gaussian function.  The mean offset of the distribution is used
to construct an unfolding bias correction to be applied to the data, while the larger of the root mean square and the Gaussian width is assigned as the statistical uncertainty 
associated with that bin in the unfolded distribution.  
The unfolding bias correction is small, generally 0.5--2\%, and always much smaller than the statistical uncertainty in the bin.  Overall, the statistical uncertainties are within 1--17\%, depending on jet multiplicity and jet $p_T$ bin.

The systematic uncertainties affecting this measurement can be divided into three types: those related to the knowledge of the detector response, those related to the background modeling and those associated with the unfolding method itself
The systematic uncertainties related to the modeling of the detector response and their effect on the final cross sections arise from the calibration of the jet energy scale [3--16\%], from the measurements of the jet energy resolution [0.1--17\%], the jet identification efficiency [0.3--4\%], the jet-track matching requirement [1--11\%], the trigger efficiency [1--4\%], the electron identification efficiency [4--5\%], and the uncertainty in the luminosity determination [6.1\%].
We determine the systematic uncertainty for all these sources apart from the latter two using the \textsc{alpgen+pythia} ensembles.  
The relevant variables in all events are varied within their systematic uncertainties, resulting in new signal templates and new migration matrices. 
The nominal ensembles (which look and behave as our reconstructed data distributions) are again unfolded but this time with inputs to \textsc{guru} replaced with the systematic-shifted samples
%(specifically, the background subtracted $W$+jets reco templates, a shifted detector response matrix and acceptance corrections).  
As expected, it is found that the statistical uncertainties from the shifted residual distributions are largely insensitive to changes in the detector response, but the unfolding bias can vary significantly.  
The change in the bias from the nominal to shifted ensembles is attributed to the systematic uncertainty in the unfolded data distributions.
%The largest contribution to the detector systematic uncertainties arises from the jet energy scale correction, yielding uncertainties in the cross section results of 3-15\% depending on the jet multiplicity and $p_T$ bin.  
%The systematic uncertainty due to the electron identification is evaluated using methods documented in~\cite{Abazov:2007kg}, and contributes less than 1\% to the final normalized result.
%The luminosity uncertainty on the differential cross sections is $\pm$6.1\% (a flat systematic on all bins). 
All differential cross section measurements are 
normalized to the measured inclusive $W$ boson cross section, resulting in a complete (partial) cancellation of the systematic uncertainties due to luminosity (trigger and electron identification efficiencies).
%%Certain systematic uncertainties on the differential cross sections contain bin-to-bin correlations.  
The dominant uncertainties due to jet energy scale and jet energy resolution are correlated bin-to-bin (and between jet spectra), 
the uncertainties due to the jet-track matching requirement and electron identification efficiency are partially correlated. All other uncertainties are considered to be uncorrelated.
The correlation of systematic uncertainties between jet multiplicity bins are taken into account when normalizing the differential cross section spectra and in determining the 
uncertainties on measurement of the $\sigma_n/\sigma_{n-1}$ inclusive cross section ratios.

The remaining sources of systematic uncertainty are the normalization and differential distributions of the multijet background [0.1--4\%], 
the uncertainty due to the electron final state radiation at particle level ($<$1\%), uncertainties associated with the unfolding method ($<$1\%) and
the theoretical uncertainty on the $t\bar{t}$ cross section.
In some regions of phase space (at high $p_T$ in the three and four jet multiplicity bins) the data sample is dominated by $t\bar{t}$ production.
In these regions the $\sim 8$\% uncertainty in the $t\bar{t}$ cross section translates into an uncertainty of up to 19\% in the $t\bar{t}$ subtracted $W$+jets signal.
Uncertainties due to the unfolding procedure come from the uncertainty on the derivation of the unfolding bias used to correct the unfolded spectra, 
and from the change of the final result when this is obtained repeating the unfolding procedure with a data-derived reweighting of the MC inputs to 
\textsc{guru} in order to account for mismodeling effects present in the Monte Carlo predictions.

%The uncertainty on the multijet background is computed in the standard way for the matrix-method~\cite{Abazov:2007kg} and contributes up to 4\% uncertainty in the cross section results at the highest $p_T$ values.  
%The systematic uncertainty associated with the $t\bar{t}$ production cross section is determined by varying the theoretical cross section within its theoretical uncertainty, scaling the $t\bar{t}$ contribution up or down by this amount, and determining a conservative limit on the associated uncertainty on the $W$+jets contribution by assuming the additional (reduced) top contribution reduces (increases) the $W$ contribution directly.  
%This implies that a reduction in the number of $t\bar{t}$ events in a given bin by N implies an increase of $W$ events in that same bin by N.

As in the case of the differential cross section measurements, the inclusive $W(\to e\nu)$+jets production cross sections are normalized to the measured inclusive $W\to e\nu$ cross section.  
This normalization reduces (or cancels) systematic uncertainties and provides sensitivity to the shape of the distribution in comparisons to Monte Carlo and theoretical predictions.  
%The inclusive $W$ cross section is corrected to the particle level by calculating: 
%$\frac{1}{\sigma_W}=\frac{\cal L}{N^{reco}_{Data}}\cdot\frac{N^{reco}_{sim}}{N^{particle}_{sim}}$
%where $\sigma_W$ is the inclusive $W$ cross section, $N^{reco}_{Data}$ is the number of inclusive $W$ events in data, $N^{reco}_{sim}$ is the number of inclusive $W$ events in reconstruction-level simulation, $N^{particle}_{sim}$ is the number of inclusive $W$ events in particle-level simulation, and ${\cal L}$ is the total integrated luminosity of the data sample.
The events passing the selection requirements are well described by the Monte Carlo predictions and the sample is dominated ($>99.8\%$) by the inclusive production of $W$ events.
The total inclusive $W$ boson cross section within the kinematic acceptance is measured to be $\sigma_W = 1097 \pm 1$(stat) $^{+39}_{-59} $(syst) $\pm~67 $(lumi)~pb.  
This number is used to normalize the differential cross section results.  

Recent theoretical work \cite{Ellis:2009bu, Berger:2009zg} has extended the availability of predictions up 
to $W$+3 jet events at NLO.  Although there has also been a recent calculation of $W$+4 jet production
at NLO for $pp$ collisions at $\sqrt{s}=7$ (or $14$)~TeV~\cite{Berger:2010zx}, these predictions are not available for the Tevatron, and comparisons with theory are therefore limited to LO for $W$+4 jet production.  In this analysis, we use the interfaced
\textsc{blackhat+sherpa}\,\cite{Berger:2009ba} and \textsc{rocket+mcfm}~\cite{Ellis:2008qc,Giele:2008bc} programs as the main sources for theoretical predictions of $W$+jets production.
The \textsc{mcfm} calculations employ version 6.0 of the program.  \textsc{blackhat} and \textsc{rocket} are parton level generators which incorporate NLO QCD calculations with up to 3 final state jets. They provide parton level jets corresponding to the hard partons, but they
 do not include the underlying event or hadronization effects.
We compare both theory predictions to our measured cross sections, in order to determine the differences that arise from theoretical choices made in the calculations, such as the choice of renormalization and factorization scales, and in order to explore the uncertainties inherent in these predictions.

The \Blackhat program employs the renormalization ($\mu_R$) and factorization ($\mu_F$) scale $\mu=\mu_F=\mu_R=\frac{1}{2}\hat{H}^\prime_T$, where $\hat{H}^\prime_T$ 
is the scalar sum of the parton and $W$ transverse energies. 
\Blackhat does not provide cross sections using the D0 midpoint jet algorithm, but instead uses the \textsc{siscone}~\cite{Salam:2007xv} algorithm with split-merge parameter $f=0.5$ and cone radius 
${\cal R} =0.5$.  In order to keep all the theory predictions on the same footing, we therefore show the \Blackhat and \textsc{rocket+mcfm} predictions using the \textsc{siscone} jet algorithm.  
The effect of differences in the theoretical predictions produced with different jet algorithms was found to be approximately one order of magnitude smaller than the scale uncertainties in all jet multiplicity bins, and so is considered to have negligible impact on the interpretation of the theory/data comparison.
%%%%However, we expect differences between the jet algorithms to be a negligible effect. 
The choice made by the \textsc{rocket+mcfm} authors is 
$\mu=\sqrt{M_W^2+\frac{1}{4}(\Sigma p^{\text{jet}})^2}$
(in the 2, 3, and 4-jet bins), summing over the four-momenta of all jets in the event, where $M_W$ is the mass of the $W$ boson.  
This scale choice was suggested in Ref.~\cite{Bauer:2009km} because it sums large logarithms in the calculation to all orders.   
In the 1-jet bin, a slightly modified choice of $\mu=\sqrt{M_W^2+ (p_T^{\text{jet}})^2}$ is used.
This is due to the fact that in the 1-jet bin, the NLO calculation includes diagrams with an extra hard (real) emission or virtual loop corrections.   
For the Born and virtual loop diagrams, the only hard scale is $M_W$, due to the single massless jet balancing the $W$ boson.  
However, in the case of diagrams with an extra hard emission, the two final state partons can be combined into one massive jet by the jet reconstruction algorithm increasing the scale of the real contributions, which generally increase the cross section.  As a result, the real diagrams are evaluated with a coupling that is smaller, due to the running of $\alpha_s$, than the virtual diagrams, which leads to a prediction of the NLO cross section that is too low.   
Both theory calculations use the MSTW2008 parton density function (PDF)~\cite{mstw}, where the LO (NLO) cross section calculation is matched to the LO (NLO) PDF.  
The uncertainties on the theory predictions are estimated by multiplying $\mu$ by factors of 2 and 0.5.

\begin{figure}[tbp]
\includegraphics[scale=0.46]{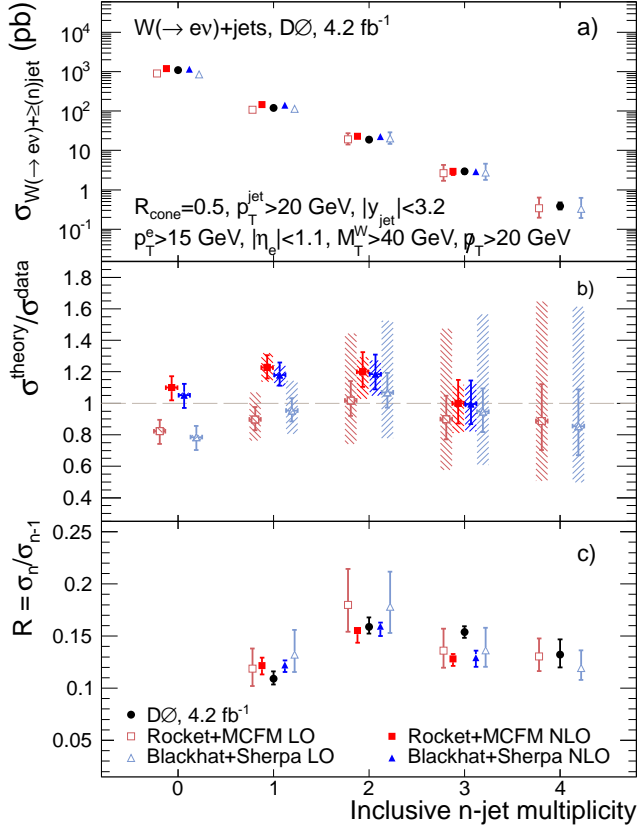}
\caption{\label{fig:inclXsec} 
(a) Total inclusive $n$-jet cross sections $\sigma_n=\sigma (W(\to e\nu)\ + \geq \hspace{-1mm} n\text{ jet;}\; p^{\text{jet}}_T>20~\textrm{GeV})$
    as a function of inclusive jet multiplicity, 
(b) the ratio of the theory predictions to the measurements, and 
(c) $\sigma_n/\sigma_{n-1}$ ratios for data, \Blackhat and \textsc{rocket+mcfm}. 
Error bars on data points represent combined statistical and systematic uncertainties on measured cross sections. The uncertainties on the theory points in (a) and (c) 
and the hashed areas in (b) represent the theoretical uncertainty arising from the choice of renormalization and factorization scale. In (b) the error bars on the points 
represent the data uncertainties.
}
\end{figure}

Fixed-order pQCD predictions provide only a parton-level prediction which is not immediately comparable to the unfolded data. 
Additional corrections must be applied to propagate the fixed order predictions to the particle level. 
The two effects which contribute to this parton-to-particle
correction are hadronization of the final state partons and the presence of the underlying event.
These corrections (referred to collectively as hadronization corrections) are obtained with 
the \sherpa MC program~\cite{Gleisberg:2008ta}, which employs the CTEQ6.6 PDF set~\cite{cteq}. The corrections are generally around 10\%, but can be as large as 25\% at high $p_T^{\text{jet}}$.  The parton level cross sections are computed with the \textsc{siscone} jet finding algorithm, while the particle level predictions are computed with the D0 midpoint cone algorithm, in order to account for the difference in jet algorithm between the data and the pQCD predictions.  The impact of folding the correction for the jet algorithm into the overall hadronization correction is small, and approximately an order of magnitude smaller than the theoretical scale uncertainties in size.  All inclusive and differential pQCD predictions
have the hadronization corrections applied to them.  We provide the tables of the hadronization corrections~\cite{appendix} so that future pQCD calculations can be compared to the data on the same terms. The quoted uncertainty on these corrections is purely statistical.

Fig.~\ref{fig:inclXsec}(a) shows the absolute inclusive $W+n$ jet cross sections for each jet multiplicity considered, compared with the LO and NLO theoretical predictions from \Blackhat and \textsc{rocket+mcfm}, where both are corrected for hadronization effects. 
Fig.~\ref{fig:inclXsec}(b) shows the ratio of theory to data.
Good agreement is observed between data and the NLO theory predictions, except for the 1-jet bin, where the NLO prediction presents a slight excess with respect to the data.
Fig.~\ref{fig:inclXsec}(c) shows the measurement of the $\sigma_n/\sigma_{n-1}$ inclusive cross section ratio as a function of inclusive jet multiplicity
for $n$=1--4 in comparison to predictions of this ratio from LO and NLO calculations.  
Here, the theoretical uncertainty takes the correlations of the scale choice between the $n$ and $n-1$ multiplicity bins into account.
The data uncertainties are also calculated from the relative uncertainties on the two cross sections, but
with partial or total cancellation of systematic uncertainties due to electron identification, trigger, and luminosity.
The uncertainties due to the jet corrections are correlated between bins and are accounted for.  
The total uncertainties on the measurement presented throughout this paper are comparable to the scale uncertainties on the predictions at NLO.
Tables of the measured and theoretical cross sections and their uncertainties are given in the appendix to this paper.

\begin{figure}[tbp]
\includegraphics[scale=0.43]{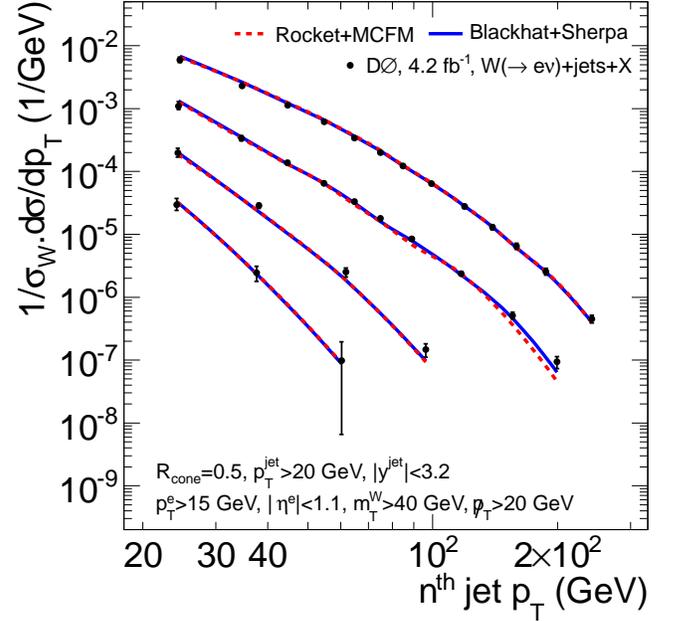}
\caption{\label{fig:finalxsecs} 
Measured $W$+n jet differential cross section as a function of $n^{\text{th}}$ jet $p_T$ for jet multiplicities $n=$1--4, normalized to the inclusive $W \rightarrow e \nu$ cross section.  
$W$+1 jet inclusive spectra are shown by the top curve, the $W$+4 jet inclusive spectra by the bottom curve.  The measurements are compared to the fixed-order NLO predictions 
for the jet multiplicities $n=$1--3, and to LO predictions for $n=4$. }
\end{figure}

The unfolded differential data cross sections (multiplied by the branching fraction of the $W\to e\nu$ decay) for each jet multiplicity are shown in Fig.~\ref{fig:finalxsecs}.   The data are normalized by the measured inclusive $W$ boson cross section in all jet multiplicity bins, which reduces the uncertainties in the measurement because of cancellation of some systematic uncertainties.  The data spectra are compared to the predictions from \textsc{rocket+mcfm} and \Blackhat (again normalized by their respective inclusive $W$ boson cross sections and corrected for hadronization effects).  The theory is able to describe the data throughout the $p_T^{\rm jet}$ spectra for all multiplicities, although a detailed comparison is best made by examining the ratios of theory to data. Each data point is placed at the $p_T$ value where the theoretical differential cross section is equal to the average cross section within the bin~\cite{WTSYDP}. 

\begin{figure}[tbp]
\includegraphics[scale=0.44]{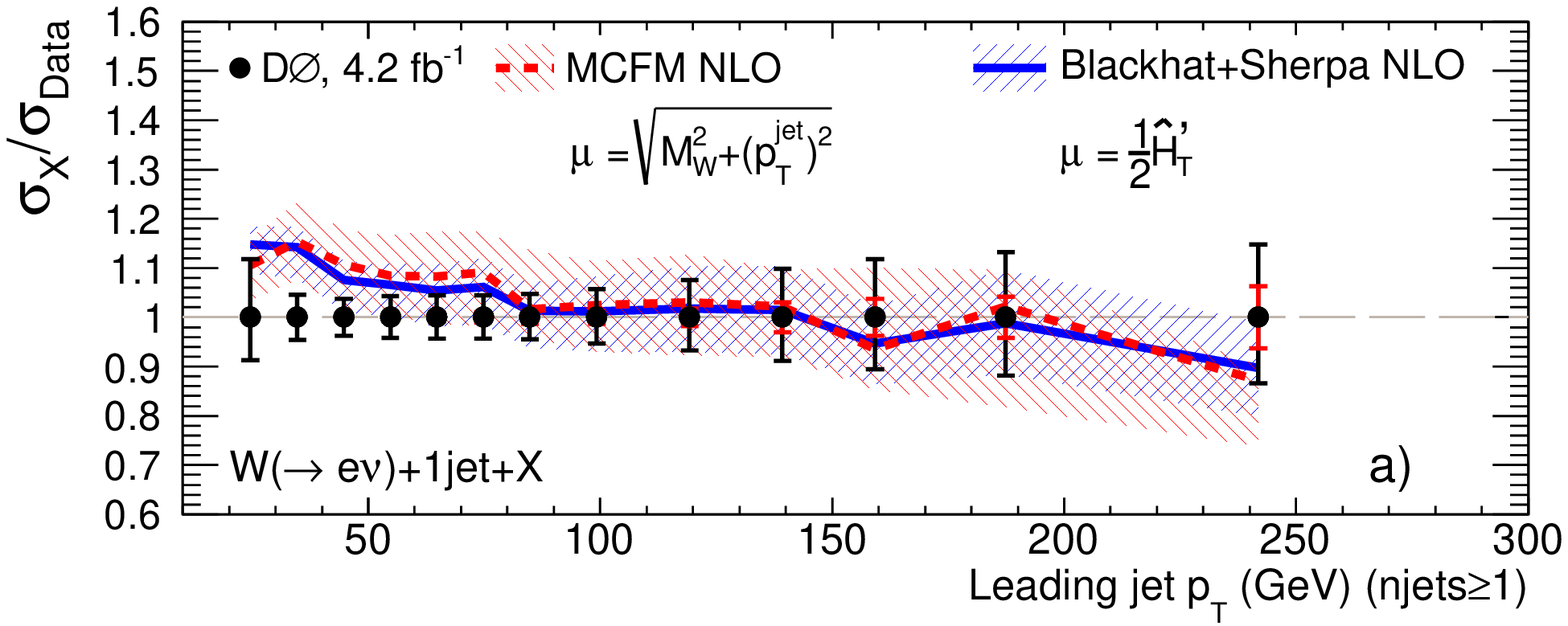}
\includegraphics[scale=0.44]{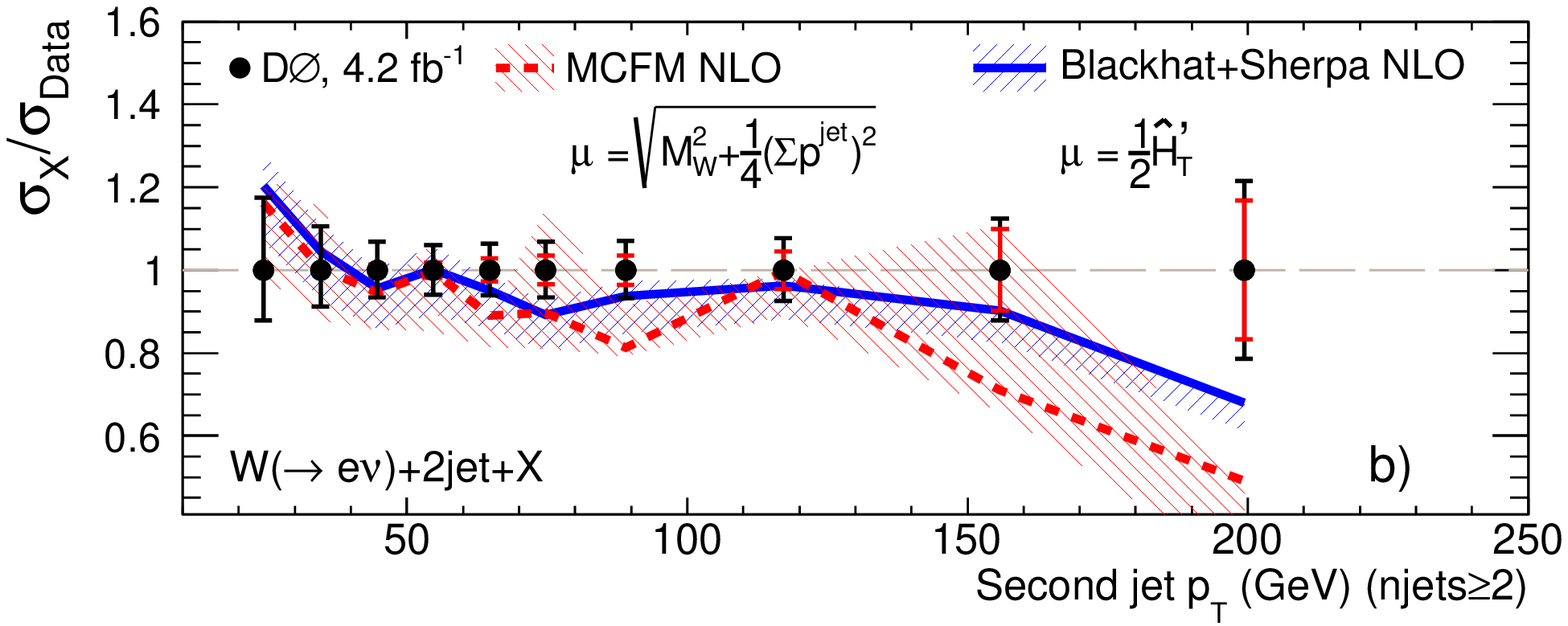}
\includegraphics[scale=0.44]{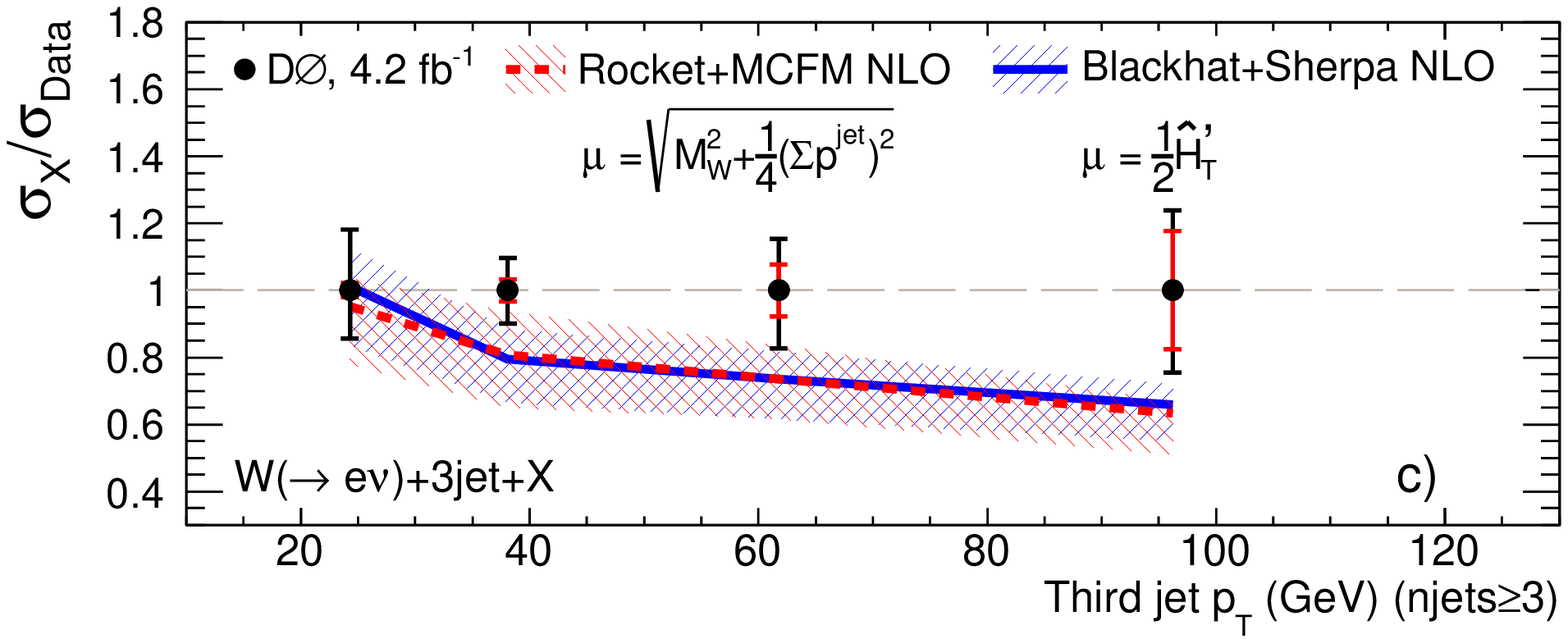}
\includegraphics[scale=0.44]{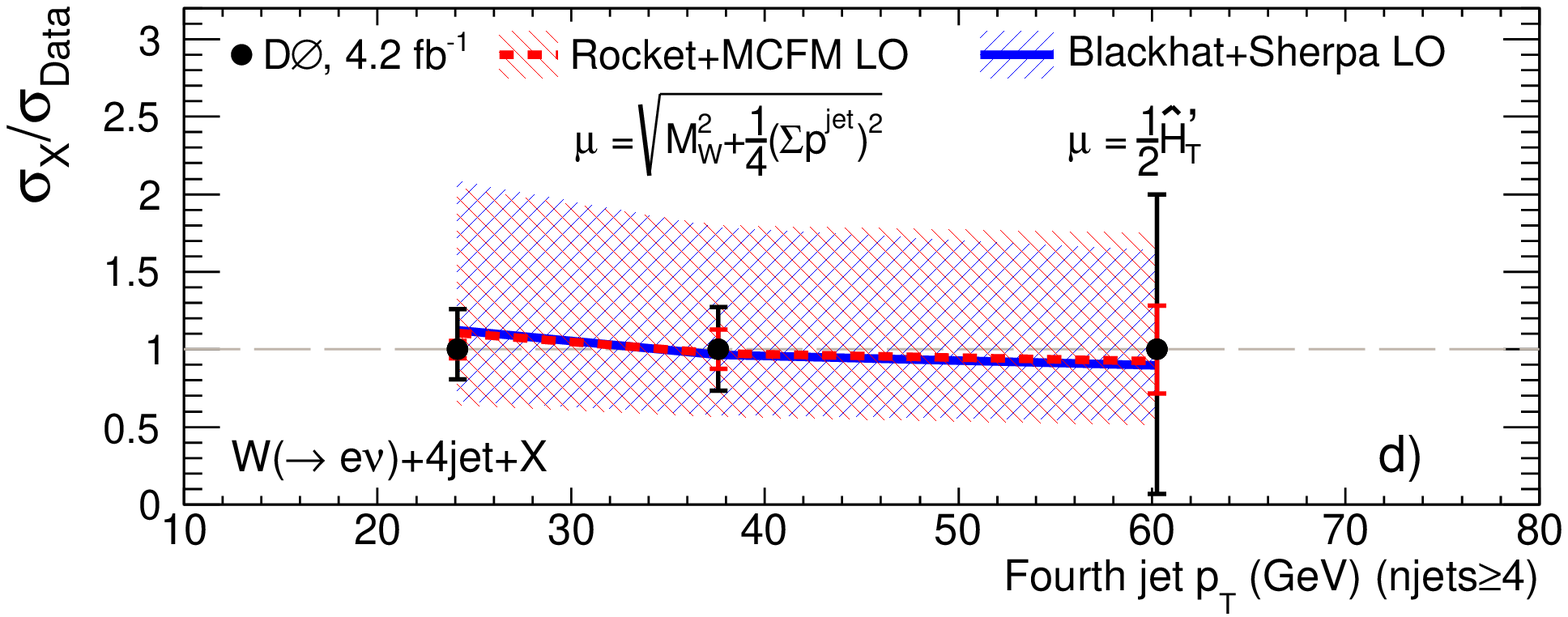}
\caption{\label{fig:ratios} The ratio of pQCD predictions to the measured differential cross sections for the $n^{\text{th}}$ jet $p_T$ in
(a) $W$+1 jet events, (b) $W$+2 jet events, (c) $W$+3 jet events, and (d) $W$+4 jet events.
%The data and theory predictions are normalized by the measured inclusive $W$ boson cross section and the predicted inclusive $W$ boson cross sections, respectively.  
The inner (red) bars represent the statistical uncertainties of the measurement, while the outer (black) bars represent the statistical and systematic uncertainties 
added in quadrature.  The shaded areas indicate the theoretical uncertainties due to variations of the factorization and renormalization scale.
}
\end{figure}

The ratio of the theory predictions to the unfolded differential data cross sections are shown in Fig.~\ref{fig:ratios}.  Each of the data and theory cross sections is 
normalized to its respective inclusive $W$ boson production cross section. In the inclusive $W$+1 jet bin [Fig.~\ref{fig:ratios}(a)], the data uncertainties vary by 4-–14\%, 
but for most jet transverse momenta these uncertainties are smaller than the theoretical uncertainties. The data agree well with both NLO theory calculations, 
although the theoretical prediction is slightly higher than the data at low $p_T^{\text{jet}}$. The inclusive $W$+2 jet bin results are shown in Fig.~\ref{fig:ratios}(b). 
The measured uncertainties vary by 5–-20\% and are similar to those of the 1-jet bin.   The \Blackhat and \textsc{rocket+mcfm} predictions are in good agreement with the data everywhere.  
In Fig.~\ref{fig:ratios}(c), the ratio of $W$+3 jet pQCD predictions to the differential cross sections are shown. The results of NLO predictions are below the data at high $p_T^{\text{jet}}$, 
but still consistent within uncertainties.  In Fig.~\ref{fig:ratios}(d), the differential cross section measurement of $W$+4 jets is shown as a ratio to the LO pQCD prediction. 
The theory prediction can reproduce the data, albeit with large uncertainties. 
Theoretical cross-sections at LO suffer from strong dependence on the choice of renormalization and factorization scales, in part due to large logarithmic corrections 
and higher order contributions. The significant reduction of the scale uncertainty at NLO compared to the same uncertainty at LO is an indication that the size of the 
NNLO corrections is small.
An NLO prediction for this final state is necessary to make a more robust comparison.

In summary, $W$+$n$ jet inclusive cross sections for $n=1, 2, 3$ and $4$ jets have been measured using 4.2 \ifb of integrated luminosity collected by the D0 detector. 
The measurements include the total inclusive cross section for each jet multiplicity and differential cross sections as a function of the $n^{\text{th}}$ jet $p_T$.  
These measurements represent a test of pQCD complementary to the extensive \Dzero $Z$+jets measurements~\cite{Abazov:2008ez,Abazov:2009av,Abazov:2009pp}.  %Moreover, $W$+jets contributes to more final states which could be sensitive to new physics, and it is important to constrain this potentially large background so that it can be modeled accurately by Monte Carlo programs used in SM Higgs and beyond the Standard Model searches.
The measured cross sections improve on the measurement by CDF~\cite{Aaltonen:2007ip} by including $W$+4 jet differential cross sections, by substantially improving 
the uncertainties on differential cross sections in all jet multiplicities, and by performing the first comparison with NLO $W$+3 jet cross section predictions. 
The measured cross sections are generally found to agree with the NLO calculation although certain regions of phase space are identified where these predictions
could better match the data.

Supplementary material including tabulated $W$+n jet cross section measurements, theoretical predictions, and hadronization corrections applied to the theory can be found 
in the appendix to this document and online at doi:10.1016/j.physletb.2011.10.011.

% acknowledgement.tex                             6 April 2011
The authors thank the \textsc{rocket+mcfm} and \textsc{blackhat+sherpa} authors for generating the theoretical predictions.   
We also thank Jan Winter for help with generating the hadronization corrections.  Many thanks go to Giulia Zanderighi, 
Fernando Febres Cordero, Lance Dixon, Zvi Bern and Jan Winter for useful discussions.

We thank the staffs at Fermilab and collaborating institutions,
and acknowledge support from the
DOE and NSF (USA);
CEA and CNRS/IN2P3 (France);
FASI, Rosatom and RFBR (Russia);
CNPq, FAPERJ, FAPESP and FUNDUNESP (Brazil);
DAE and DST (India);
Colciencias (Colombia);
CONACyT (Mexico);
KRF and KOSEF (Korea);
CONICET and UBACyT (Argentina);
FOM (The Netherlands);
STFC and the Royal Society (United Kingdom);
MSMT and GACR (Czech Republic);
CRC Program and NSERC (Canada);
BMBF and DFG (Germany);
SFI (Ireland);
The Swedish Research Council (Sweden);
and
CAS and CNSF (China).
%
   % input acknowledgement

\clearpage
\pagebreak
\newpage

\begin{widetext}

{\Large Appendix: Tables of Measurements, pQCD calculations and non-perturbative hadronization corrections}
\vspace{0.5cm}

In this appendix, we provide tables of the measured differential cross sections, theory predictions and hadronization corrections described in this paper.   The region that defines the kinematic phase space of the measurement at particle level is given by the electron transverse momentum, $p_T^e \ge 15$ GeV, and pseudorapidity $|\eta^e| < $ 1.1, total transverse energy of all neutrinos $E_T^{\nu} > 20$ GeV, $W$ transverse mass $M_T^W > 40$ GeV, jet transverse momentum $p_T^{\rm jet} \ge$ 20~GeV and rapidity $|y^{\rm jet}| < $ 3.2.  
$W$ boson inclusive cross sections per jet multiplicity bin correspond to the sum over all $p_T^{\rm jet}$ in the given jet multiplicity, and the normalized cross sections are the absolute cross sections in a given jet multiplicity divided by the inclusive $W$ boson cross sections in the entire kinematic region.  The hadronization corrections can be applied as a multiplicative factor to parton level jets clustered using the \textsc{siscone} algorithm to produce particle jets, as defined by the D0 midpoint algorithm. Quoted systematic uncertainties on measured {\em absolute} cross-sections include a 6.1\% luminosity uncertainty.
%\section{JET 1 PT}

\begin{table}[tbfh]
\caption{Measured differential cross section, normalized to the measured inclusive $W$ cross section, as a function of leading jet $p_T$ for events with one or more identified jets,
 along with statistical and systematic uncertainties.} 
\label{tb:diffxsec_jet_pt_1}
\begin{ruledtabular}
\begin{tabular}{l l r}
$E^{\textrm{\tiny Leading jet}}_{T}$ (GeV) & $\langle E^{\textrm{\tiny Leading jet}}_{T}\rangle$ (GeV) & {$1/\sigma_{W} \cdot d\sigma/dp^{\textrm{\tiny Leading jet}}_{T}$ (1/GeV)} \\
\hline
$20-30$ & $ 24.6$ & $\big( 5.95\pm 0.02\textrm{(stat)} {}^{+0.71}_{-0.52} \textrm{(syst)} \big) \times 10^{-3}$ \\
$30-40$ & $ 34.7$ & $\big( 2.30\pm 0.01\textrm{(stat)} {}^{+0.10}_{-0.11} \textrm{(syst)} \big) \times 10^{-3}$ \\
$40-50$ & $ 44.7$ & $\big( 1.13\pm 0.01\textrm{(stat)} \pm 0.04 \textrm{(syst)} \big) \times 10^{-3}$ \\
$50-60$ & $ 54.8$ & $\big( 6.16\pm 0.05\textrm{(stat)} \pm 0.26 \textrm{(syst)} \big) \times 10^{-4}$ \\
$60-70$ & $ 64.8$ & $\big( 3.43\pm 0.03\textrm{(stat)}  {}^{+0.15}_{-0.14} \textrm{(syst)} \big) \times 10^{-4}$ \\
$70-80$ & $ 74.8$ & $\big( 2.00\pm 0.02\textrm{(stat)}  {}^{+0.09}_{-0.08} \textrm{(syst)} \big) \times 10^{-4}$ \\
$80-90$ & $ 84.8$ & $\big( 1.22\pm 0.02\textrm{(stat)}  {}^{+0.06}_{-0.05} \textrm{(syst)} \big) \times 10^{-4}$ \\
$90-110$ & $ 99.3$ & $\big( 6.43\pm 0.09\textrm{(stat)}  {}^{+0.36}_{-0.33} \textrm{(syst)} \big) \times 10^{-5}$ \\
$110-130$ & $ 119.3$ & $\big( 2.76\pm 0.05\textrm{(stat)}  {}^{+0.20}_{-0.18} \textrm{(syst)} \big) \times 10^{-5}$ \\
$130-150$ & $ 139.3$ & $\big( 1.29\pm 0.04\textrm{(stat)}  {}^{+0.12}_{-0.11} \textrm{(syst)} \big) \times 10^{-5}$ \\
$150-170$ & $ 159.2$ & $\big( 6.45\pm 0.24\textrm{(stat)}  {}^{+0.72}_{-0.64} \textrm{(syst)} \big) \times 10^{-6}$ \\
$170-210$ & $ 187.4$ & $\big( 2.53\pm 0.10\textrm{(stat)}  {}^{+0.32}_{-0.28} \textrm{(syst)} \big) \times 10^{-6}$ \\
$210-300$ & $ 241.8$ & $\big( 4.48\pm 0.28\textrm{(stat)}  {}^{+0.60}_{-0.53} \textrm{(syst)} \big) \times 10^{-7}$ \\
\hline\hline
\multicolumn{2}{l}{(Normalized) cross section:} & $0.109 \pm 0.0002 \textrm{(stat)} {}^{+0.007}_{-0.005} \textrm{(syst)}$ \\
\end{tabular}
\end{ruledtabular}
\end{table}

\begin{table}[tbfh]
\caption{Normalized NLO theory predictions for $W$+1 jet cross sections before application of hadronization corrections for the leading jet $p_T$.}
\label{tb:XXX}
\begin{ruledtabular}
\begin{tabular}{l l l r}
$E^{\textrm{\tiny Leading jet}}_{T}$ (GeV) & {$1/\sigma_{W} \cdot d\sigma/dp^{\textrm{\tiny Leading jet}}_{T}$ (1/GeV)} & $+1$ standard deviation & $-1$ standard deviation\\
\hline
\multicolumn{4}{c}{\textsc{mcfm} cross section predictions normalized by the \textsc{mcfm} inclusive $W$ boson cross section} \\
\hline
20-30  &  $5.91\times 10^{-3} $&$ +3.0\times 10^{-4} $&$ -3.7\times 10^{-4}$\\
30-40  &  $2.74\times 10^{-3} $&$ +1.9\times 10^{-4} $&$ -1.8\times 10^{-4}$\\
40-50  &  $1.38\times 10^{-3} $&$ +0.9\times 10^{-4} $&$ -1.1\times 10^{-4}$\\
50-60  &  $7.30\times 10^{-4} $&$ +5.5\times 10^{-5} $&$ -5.4\times 10^{-5}$\\
60-70  &  $4.16\times 10^{-4} $&$ +3.5\times 10^{-5} $&$ -3.7\times 10^{-5}$\\
70-80  &  $2.44\times 10^{-4} $&$ +1.8\times 10^{-5} $&$ -2.4\times 10^{-5}$\\
80-90  &  $1.44\times 10^{-4} $&$ +1.7\times 10^{-5} $&$ -0.8\times 10^{-5}$\\
90-110  &  $7.59\times 10^{-5} $&$ +6.6\times 10^{-6} $&$ -6.9\times 10^{-6}$\\
110-130  &  $3.35\times 10^{-5} $&$ +3.2\times 10^{-6} $&$ -3.5\times 10^{-6}$\\
130-150  &  $1.56\times 10^{-5} $&$ +1.1\times 10^{-6} $&$ -1.6\times 10^{-6}$\\
150-170  &  $7.34\times 10^{-6} $&$ +1.3\times 10^{-6} $&$ -0.6\times 10^{-6}$\\
170-210  &  $3.28\times 10^{-6} $&$ +2.4\times 10^{-7} $&$ -6.6\times 10^{-7}$\\
210-300  &  $5.30\times 10^{-7} $&$ +1.9\times 10^{-8} $&$ -8.0\times 10^{-8}$\\
\hline
\multicolumn{4}{c}{\textsc{blackhat+sherpa} cross section predictions normalized by the \textsc{blackhat+sherpa} inclusive $W$ boson cross section} \\
\hline
20-30  &  $6.14\times 10^{-3} $&$ +1.9\times 10^{-4} $&$ -3.1\times 10^{-4}$\\
30-40  &  $2.71\times 10^{-3} $&$ +1.0\times 10^{-4} $&$ -1.5\times 10^{-4}$\\
40-50  &  $1.34\times 10^{-3} $&$ +4.7\times 10^{-5} $&$ -7.7\times 10^{-5}$\\
50-60  &  $7.18\times 10^{-4} $&$ +2.9\times 10^{-5} $&$ -4.5\times 10^{-5}$\\
60-70  &  $4.05\times 10^{-4} $&$ +2.1\times 10^{-5} $&$ -2.8\times 10^{-5}$\\
70-80  &  $2.37\times 10^{-4} $&$ +1.3\times 10^{-5} $&$ -1.7\times 10^{-5}$\\
80-90  &  $1.43\times 10^{-4} $&$ +0.8\times 10^{-5} $&$ -1.0\times 10^{-5}$\\
90-110  &  $7.49\times 10^{-5} $&$ +5.3\times 10^{-6} $&$ -6.0\times 10^{-6}$\\
110-130  &  $3.31\times 10^{-5} $&$ +2.6\times 10^{-6} $&$ -2.9\times 10^{-6}$\\
130-150  &  $1.55\times 10^{-5} $&$ +1.3\times 10^{-6} $&$ -1.4\times 10^{-6}$\\
150-170  &  $7.43\times 10^{-6} $&$ +6.0\times 10^{-7} $&$ -6.4\times 10^{-7}$\\
170-210  &  $3.17\times 10^{-6} $&$ +3.4\times 10^{-7} $&$ -3.3\times 10^{-7}$\\
210-300  &  $5.46\times 10^{-7} $&$ +5.6\times 10^{-8} $&$ -6.0\times 10^{-8}$\\
\end{tabular}
\end{ruledtabular}
\end{table}

\begin{table}[tbfh]
\caption{Hadronization corrections derived with \textsc{sherpa} 1.2.3 and the CTEQ6.6 PDF set for the leading jet $p_T$ distribution.}
\label{tb:XXX}
\begin{ruledtabular}
\begin{tabular}{l r}
%$E^{\textrm{\tiny Leading jet}}_{T}$ (GeV) & \multicolumn{2}{c}{Hadronization correction} \\
$E^{\textrm{\tiny Leading jet}}_{T}$ (GeV) & Hadronization correction \\
\hline
20-30 & 1.110 $\pm$ 0.004\\
30-40 & 0.968 $\pm$ 0.005\\
40-50 & 0.907 $\pm$ 0.005\\
50-60 & 0.914 $\pm$ 0.007\\
60-70 & 0.892 $\pm$ 0.008\\
70-80 & 0.898 $\pm$ 0.009\\
80-90 & 0.87 $\pm$ 0.01\\
90-110 & 0.87 $\pm$ 0.01\\
110-130 & 0.85 $\pm$ 0.01\\
130-150 & 0.85 $\pm$ 0.02\\
150-170 & 0.82 $\pm$ 0.02\\
170-210 & 0.79 $\pm$ 0.02\\
210-300 & 0.74 $\pm$ 0.03\\
\end{tabular}
\end{ruledtabular}
\end{table}

%\section{JET 2 PT}

\begin{table}[tbfh]
\caption{Measured differential cross section, normalized to the measured inclusive $W$ cross section, as a function of second jet $p_T$ for events with two or more identified jets,
 along with statistical and systematic uncertainties.} 
\label{tb:diffxsec_jet_pt_2}
\begin{ruledtabular}
\begin{tabular}{l l r}
$E^{\textrm{\tiny Second jet}}_{T}$ (GeV) & $\langle E^{\textrm{\tiny Second jet}}_{T}\rangle$ (GeV) & $1/\sigma_{W} \cdot d\sigma/dp^{\textrm{\tiny Second jet}}_{T}$ (1/GeV) \\
\hline
$20-30$ & $ 24.4$ & $\big( 1.10\pm 0.01 \textrm{(stat)} {}^{+0.19}_{-0.13} \textrm{(syst)} \big) \times 10^{-3}$ \\
$30-40$ & $ 34.6$ & $\big( 3.38\pm 0.03 \textrm{(stat)} {}^{+0.36}_{-0.30} \textrm{(syst)} \big) \times 10^{-4}$ \\
$40-50$ & $ 44.7$ & $\big( 1.39\pm 0.02 \textrm{(stat)} {}^{+0.09}_{-0.09} \textrm{(syst)} \big) \times 10^{-4}$ \\
$50-60$ & $ 54.7$ & $\big( 6.53\pm 0.12 \textrm{(stat)} \pm 0.37 \textrm{(syst)} \big) \times 10^{-5}$ \\
$60-70$ & $ 64.7$ & $\big( 3.33\pm 0.09 \textrm{(stat)} {}^{+0.19}_{-0.18} \textrm{(syst)} \big) \times 10^{-5}$ \\
$70-80$ & $ 74.8$ & $\big( 1.80\pm 0.06 \textrm{(stat)} {}^{+0.11}_{-0.10} \textrm{(syst)} \big) \times 10^{-5}$ \\
$80-100$ & $ 89.0$ & $\big( 8.51\pm 0.30 \textrm{(stat)} {}^{+0.52}_{-0.50} \textrm{(syst)} \big) \times 10^{-6}$ \\
$100-140$ & $ 117.1$ & $\big( 2.38\pm 0.11 \textrm{(stat)} {}^{+0.15}_{-0.14} \textrm{(syst)} \big) \times 10^{-6}$ \\
$140-180$ & $ 155.8$ & $\big( 5.15\pm 0.51 \textrm{(stat)} {}^{+0.40}_{-0.37} \textrm{(syst)} \big) \times 10^{-7}$ \\
$180-250$ & $ 199.3$ & $\big( 9.40\pm 1.6 \textrm{(stat)} {}^{+1.3}_{-1.2} \textrm{(syst)} \big) \times 10^{-8}$ \\
\hline\hline
\multicolumn{2}{l}{(Normalized) cross section:} & $0.017 \pm 0.0001 \textrm{(stat)} {}^{+2.0\times 10^{-3}}_{-1.4\times 10^{-3}} \textrm{(syst)}$ \\
\end{tabular}
\end{ruledtabular}
\end{table}

\begin{table}[tbfh]
\caption{Normalized NLO theory predictions for $W$+2 jet cross sections before application of hadronization corrections for the second jet $p_T$.}
\label{tb:XXX}
\begin{ruledtabular}
\begin{tabular}{l l l r}
$E^{\textrm{\tiny Second jet}}_{T}$ (GeV) & $1/\sigma_{W} \cdot d\sigma/dp^{\textrm{\tiny Second jet}}_{T}$ (1/GeV) & $+1$ standard deviation & $-1$ standard deviation\\
\hline
\multicolumn{4}{c}{\textsc{mcfm} cross section predictions normalized by the \textsc{mcfm} inclusive $W$ boson cross section} \\
\hline
20-30  &  $1.1\times 10^{-3} $&$ +6.2\times 10^{-5}  $&$ -7.3\times 10^{-5}$\\
30-40  &  $3.6\times 10^{-4} $&$ +5.1\times 10^{-5}  $&$ -1.5\times 10^{-5}$\\
40-50  &  $1.5\times 10^{-4} $&$ +0.0 $&$ -1.4\times 10^{-5}$ \\
50-60  &  $7.1\times 10^{-5} $&$ +4.9\times 10^{-6} $&$ -5.7\times 10^{-6}$ \\
60-70  &  $3.3\times 10^{-5} $&$ +0.0 $&$ -2.8\times 10^{-6}$ \\
70-80  &  $2.0\times 10^{-5} $&$ +5.4\times 10^{-6} $&$ -1.7\times 10^{-6}$ \\
80-100  &  $8.1\times 10^{-6} $&$ +1.0\times 10^{-6} $&$ -8.1\times 10^{-7}$ \\
100-140  &  $2.7\times 10^{-6} $&$ +0.0 $&$ -2.9\times 10^{-7}$ \\
140-180  &  $4.3\times 10^{-7} $&$ +2.4\times 10^{-7} $&$ -3.9\times 10^{-8}$\\
180-250  &  $7.5\times 10^{-8} $&$ +0.0 $&$ -3.4\times 10^{-9}$\\
\hline
\multicolumn{4}{c}{\textsc{blackhat+sherpa} cross section predictions normalized by the \textsc{blackhat+sherpa} inclusive $W$ boson cross section} \\
\hline
20-30  &  $1.2\times 10^{-3} $&$ +0.7\times 10^{-4} $&$ -1.3\times 10^{-4}$\\
30-40  &  $3.7\times 10^{-4} $&$ +2.1\times 10^{-5} $&$ -4.1\times 10^{-5}$\\
40-50  &  $1.5\times 10^{-4} $&$ +0.7\times 10^{-5} $&$ -1.5\times 10^{-5}$\\
50-60  &  $7.1\times 10^{-5} $&$ +3.9\times 10^{-6} $&$ -7.6\times 10^{-6}$\\
60-70  &  $3.6\times 10^{-5} $&$ +1.2\times 10^{-6} $&$ -3.5\times 10^{-6}$\\
70-80  &  $2.0\times 10^{-5} $&$ +0.8\times 10^{-6} $&$ -1.8\times 10^{-6}$\\
80-100  &  $9.4\times 10^{-6} $&$ +3.5\times 10^{-7} $&$ -9.1\times 10^{-7}$\\
100-140  &  $2.6\times 10^{-6} $&$ +0.4\times 10^{-7} $&$ -2.2\times 10^{-7}$\\
140-180  &  $5.5\times 10^{-7} $&$ +0.0 $&$ -4.6\times 10^{-8}$\\
180-250  &  $1.0\times 10^{-7} $&$ +0.0 $&$ -1.0\times 10^{-8}$\\
\end{tabular}
\end{ruledtabular}
\end{table}

\begin{table}[tbfh]
\caption{Hadronization corrections derived with \textsc{sherpa} 1.2.3 and CTEQ6.6 PDF set for the second jet $p_T$ distribution.}
\label{tb:XXX}
\begin{ruledtabular}
\begin{tabular}{l r}
$E^{\textrm{\tiny Second jet}}_{T}$ (GeV) & Hadronization correction \\
\hline
20-30 & 1.133 $\pm$ 0.006\\
30-40 & 0.942 $\pm$ 0.009\\
40-50 & 0.89 $\pm$ 0.01\\
50-60 & 0.92 $\pm$ 0.02\\
60-70 & 0.88 $\pm$ 0.02\\
70-80 & 0.82 $\pm$ 0.03\\
80-100 & 0.85 $\pm$ 0.03\\
100-140 & 0.88 $\pm$ 0.03\\
140-180 & 0.85 $\pm$ 0.05\\
180-250 & 0.62 $\pm$ 0.05\\
\end{tabular}
\end{ruledtabular}
\end{table}

%\section{JET 3 PT}

\begin{table}[tbfh]
\caption{Measured differential cross section, normalized to the measured inclusive $W$ cross section, as a function of third jet $p_T$ for events with three or more identified jets,
 along with statistical and systematic uncertainties.} 
\label{tb:diffxsec_jet_pt_3}
\begin{ruledtabular}
\begin{tabular}{l l r}
$E^{\textrm{\tiny Third jet}}_{T}$ (GeV) & $\langle E^{\textrm{\tiny Third jet}}_{T}\rangle$ (GeV) & $1/\sigma_{W} \cdot d\sigma/dp^{\textrm{\tiny Third jet}}_{T}$ (1/GeV) \\
\hline
$20-30$ & $ 24.3$ & $\big( 2.00\pm 0.04 \textrm{(stat)} {}^{+0.36}_{-0.28} \textrm{(syst)} \big) \times 10^{-4} $ \\
$30-50$ & $ 38.1$ & $\big( 2.89\pm 0.10 \textrm{(stat)} {}^{+0.26}_{-0.27} \textrm{(syst)} \big) \times 10^{-5} $ \\
$50-80$ & $ 61.8$ & $\big( 2.54\pm 0.20 \textrm{(stat)} {}^{+0.34}_{-0.39} \textrm{(syst)} \big) \times 10^{-6} $ \\
$80-130$ & $ 96.2$ & $\big( 1.48\pm 0.26 \textrm{(stat)} {}^{+0.24}_{-0.25} \textrm{(syst)} \big) \times 10^{-7} $ \\
\hline\hline
\multicolumn{2}{l}{(Normalized) cross section:} & $0.0026 \pm 0.5\times 10^{-4} \textrm{(stat)} {}^{+3.6\times 10^{-4}}_{-2.9\times 10^{-4}} \textrm{(syst)}$ \\
\end{tabular}
\end{ruledtabular}
\end{table}

\begin{table}[tbfh]
\caption{Normalized NLO theory predictions before application of hadronization corrections for third jet $p_T$.}
\label{tb:XXX}
\begin{ruledtabular}
\begin{tabular}{l l l r}
$E^{\textrm{\tiny Third jet}}_{T}$ (GeV) & $1/\sigma_{W} \cdot d\sigma/dp^{\textrm{\tiny Third jet}}_{T}$ (1/GeV) & $+1$ standard deviation & $-1$ standard deviation\\
\hline
\multicolumn{4}{c}{\textsc{rocket+mcfm} cross section predictions normalized by the \textsc{rocket+mcfm} inclusive $W$ boson cross section} \\
\hline
20-30  &  $1.7\times 10^{-4} $&$ +5.3\times 10^{-6} $&$ -3.1\times 10^{-5}$\\
30-50  &  $2.7\times 10^{-5} $&$ +4.2\times 10^{-6} $&$ -5.2\times 10^{-6}$\\
50-80  &  $2.2\times 10^{-6} $&$ +3.0\times 10^{-7} $&$ -3.6\times 10^{-7}$\\
80-130  &  $1.2\times 10^{-7} $&$ +0.7\times 10^{-8} $&$ -2.4\times 10^{-8}$\\
\hline
\multicolumn{4}{c}{\textsc{blackhat+sherpa} cross section predictions normalized by the \textsc{blackhat+sherpa} inclusive $W$ boson cross section} \\
\hline
20-30  &  $1.8\times 10^{-4} $&$ +2.1\times 10^{-5} $&$ -3.2\times 10^{-5}$\\
30-50  &  $2.6\times 10^{-5} $&$ +3.0\times 10^{-6} $&$ -4.3\times 10^{-6}$\\
50-80  &  $2.2\times 10^{-6} $&$ +2.4\times 10^{-7} $&$ -3.6\times 10^{-7}$\\
80-130  &  $1.3\times 10^{-7} $&$ +1.2\times 10^{-8} $&$ -2.1\times 10^{-8}$\\
\end{tabular}
\end{ruledtabular}
\end{table}

\begin{table}[tbfh]
\caption{Hadronization corrections derived with \textsc{sherpa} 1.2.3 and the CTEQ6.6 PDF set for the third jet $p_T$ distribution.}
\label{tb:XXX}
\begin{ruledtabular}
\begin{tabular}{l r}
$E^{\textrm{\tiny Third jet}}_{T}$ (GeV) & Hadronization correction \\
\hline
20-30 & 1.09 $\pm$ 0.01\\
30-50 & 0.87 $\pm$ 0.01\\
50-80 & 0.83 $\pm$ 0.02\\
80-130 & 0.75 $\pm$ 0.03\\
\end{tabular}
\end{ruledtabular}
\end{table}

%\section{JET 4 PT}
\clearpage
\newpage

\begin{table}[tbfh]
\caption{Measured differential cross section, normalized to the measured inclusive $W$ cross section, as a function of fourth jet $p_T$ for events with four or more identified jets,
 along with statistical and systematic uncertainties.} 
\label{tb:diffxsec_jet_pt_4}
\begin{ruledtabular}
\begin{tabular}{l l r}
$E^{\textrm{\tiny Fourth jet}}_{T}$ (GeV) & $\langle E^{\textrm{\tiny Fourth jet}}_{T}\rangle$ (GeV) & $1/\sigma_{W} \cdot d\sigma/dp^{\textrm{\tiny Fourth jet}}_{T}$ (1/GeV) \\
\hline
$20-30$ & $ 24.1$ & $\big( 3.00\pm 0.17 \textrm{(stat)} {}^{+0.76}_{-0.55} \textrm{(syst)} \big) \times 10^{-5}$ \\
$30-50$ & $ 37.6$ & $\big( 2.46\pm 0.31 \textrm{(stat)} {}^{+0.59}_{-0.57} \textrm{(syst)} \big) \times 10^{-6}$ \\
$50-80$ & $ 60.3$ & $\big( 9.87\pm 2.78 \textrm{(stat)} {}^{+9.45}_{-8.77} \textrm{(syst)} \big) \times 10^{-8}$ \\
\hline\hline
\multicolumn{2}{l}{(Normalized) cross section:} & $0.00035 \pm 1.8\times 10^{-5} \textrm{(stat)} {}^{+7.6\times 10^{-5}}_{-5.6\times 10^{-5}} \textrm{(syst)}$ \\
\end{tabular}
\end{ruledtabular}
\end{table}

\begin{table}[tbfh]
\caption{Normalized LO theory predictions before application of hadronization corrections for fourth jet $p_T$.}
\label{tb:yyy}
\begin{ruledtabular}
\begin{tabular}{l l l r}
$E^{\textrm{\tiny Fourth jet}}_{T}$ (GeV) & $1/\sigma_{W} \cdot d\sigma/dp^{\textrm{\tiny Fourth jet}}_{T}$ (1/GeV) & $+1$ standard deviation & $-1$ standard deviation\\
\hline
\multicolumn{4}{c}{\textsc{rocket+mcfm} cross section predictions normalized by the \textsc{rocket+mcfm} inclusive $W$ boson cross section} \\
\hline
20-30  &  $2.9\times 10^{-5} $&$ +2.5\times 10^{-5} $&$ -1.2\times 10^{-5}$\\
30-50  &  $2.7\times 10^{-6} $&$ +2.3\times 10^{-6} $&$ -1.1\times 10^{-6}$\\
50-80  &  $1.2\times 10^{-7} $&$ +1.1\times 10^{-7} $&$ -5.4\times 10^{-8}$\\
\hline
\multicolumn{4}{c}{\textsc{blackhat+sherpa} cross section predictions normalized by the \textsc{blackhat+sherpa} inclusive $W$ boson cross section} \\
\hline
20-30  &  $3.0\times 10^{-5} $&$ +2.5\times 10^{-5} $&$ -1.2\times 10^{-5}$\\
30-50  &  $2.7\times 10^{-6} $&$ +2.3\times 10^{-6} $&$ -1.1\times 10^{-6}$\\
50-80  &  $1.2\times 10^{-7} $&$ +9.8\times 10^{-8} $&$ -4.7\times 10^{-8}$\\
\end{tabular}
\end{ruledtabular}
\end{table}

\begin{table}[tbfh]
\caption{Hadronization corrections derived with \textsc{sherpa} 1.2.3 and the CTEQ6.6 PDF set for the fourth jet $p_T$ distribution.}
\label{tb:XXX}
\begin{ruledtabular}
\begin{tabular}{l r}
$E^{\textrm{\tiny Fourth jet}}_{T}$ (GeV) & Hadronization correction \\
\hline
20-30 & 1.12 $\pm$ 0.02\\
30-50 & 0.86 $\pm$ 0.02\\
50-80 & 0.74 $\pm$ 0.03\\
\end{tabular}
\end{ruledtabular}
\end{table}

\clearpage
\newpage

%\section{INCLUSIVE CROSS SECTIONS}

\begin{table}[tbfh]
\caption{Measurement of $n^{\text{th}}$ jet inclusive cross section in association with a $W$ boson.  All results are given in pb.}
\label{tb:XXX}
\begin{ruledtabular}
\begin{tabular}{l r r r}
Bin & $\sigma_{W+\textrm{n-jet}}$ (pb) & Statistical Uncertainty (pb) & Systematic Uncertainty (pb) \\
\hline
$W+0$-jet & 1097 &$\pm$ 0.8 (stat) & ${}^{+79}_{-89}$ (syst)\\
$W+1$-jet & 119.5  &$\pm$ 0.3 (stat) & ${}^{+9.4}_{-8.3}$ (syst)\\
$W+2$-jet & 19.0 &$\pm$ 0.1 (stat) & ${}^{+2.4}_{-1.9}$ (syst)\\
$W+3$-jet & 2.9  &$\pm$ 0.1 (stat) & ${}^{+0.4}_{-0.4}$ (syst)\\
$W+4$-jet & 0.39  &$\pm$ 0.02 (stat)& ${}^{+0.09}_{-0.07}$ (syst)\\
\end{tabular}
\end{ruledtabular}
\end{table}

\begin{table}[tbfh]
\caption{\textsc{blackhat+sherpa} and \textsc{rocket+mcfm} predictions of $n^{\text{th}}$ jet inclusive cross sections in association with a $W$ boson.  All results are given before hadronization corrections.}
\label{tb:XXX}
\begin{ruledtabular}
\begin{tabular}{l r r}
Bin & $\sigma_{W+\textrm{n-jet}}$ (pb) & Scale Uncertainty (pb) \\
\hline
 & {\textsc{blackhat+sherpa}, no hadronization correction} \\
\hline
$W+0$-jet & 1153 & ${}^{+17}_{-7}$ (syst)\\
$W+1$-jet & 138.1  & ${}^{+7.1}_{-8.4}$ (syst)\\
$W+2$-jet & 21.4  & ${}^{+1.6}_{-2.5}$ (syst)\\
$W+3$-jet & 2.81  & ${}^{+0.36}_{-0.50}$ (syst)\\
$W+4$-jet & 0.31  & ${}^{+0.27}_{-0.13}$ (syst)\\
\hline
 & {\textsc{rocket+mcfm}, no hadronization correction} \\
\hline
$W+0$-jet & 1207 & ${}^{+13}_{-6}$ (syst)\\
$W+1$-jet & 143.6  & ${}^{+10.6}_{-10.6}$ (syst)\\
$W+2$-jet & 21.6  & ${}^{+1.7}_{-3.1}$ (syst)\\
$W+3$-jet & 2.82  & ${}^{+0.33}_{-0.53}$ (syst)\\
$W+4$-jet & 0.32  & ${}^{+0.27}_{-0.14}$ (syst)\\
\end{tabular}
\end{ruledtabular}
\end{table}

\clearpage

\begin{table}[tbfh]
\caption{Measurement of $n^{\text{th}}$ jet to $(n-1)^{\text{th}}$ jet inclusive cross section ratios.}
\label{tb:XXX}
\begin{ruledtabular}
\begin{tabular}{l r}
$\sigma_{n}/\sigma_{n-1}$ & {Ratio (data)} \\
\hline
$\sigma_{1}/\sigma_{0}$ & 0.109${}^{+0.007}_{-0.005}$ \\
$\sigma_{2}/\sigma_{1}$ & 0.159${}^{+0.009}_{-0.006}$ \\
$\sigma_{3}/\sigma_{2}$ & 0.154${}^{+0.005}_{-0.006}$ \\
$\sigma_{4}/\sigma_{3}$ & 0.132${}^{+0.015}_{-0.012}$ \\
\end{tabular}
\end{ruledtabular}
\end{table}

\begin{table}[tbfh]
\caption{\textsc{blackhat+sherpa} and \textsc{rocket+mcfm} predictions of $n^{\text{th}}$ jet to $(n-1)^{\text{th}}$ jet inclusive cross section ratios before hadronization corrections.}
\label{tb:XXX}
\begin{ruledtabular}
\begin{tabular}{l r}
$\sigma_{n}/\sigma_{n-1}$ & {Ratio (\textsc{blackhat+sherpa}, no hadronization correction)} \\
\hline
$\sigma_{1}/\sigma_{0}$ & 0.120${}^{+0.004}_{-0.007}$\\
$\sigma_{2}/\sigma_{1}$ & 0.155${}^{+0.003}_{-0.009}$\\
$\sigma_{3}/\sigma_{2}$ & 0.131${}^{+0.007}_{-0.009}$\\
$\sigma_{4}/\sigma_{3}$ & 0.11${}^{+0.02}_{-0.01}$\\
\hline
$\sigma_{n}/\sigma_{n-1}$ & {Ratio (\textsc{rocket+mcfm}, no hadronization correction)} \\
\hline
$\sigma_{1}/\sigma_{0}$ & 0.119${}^{+0.008}_{-0.008}$\\
$\sigma_{2}/\sigma_{1}$ & 0.151${}^{+0.0005}_{-0.011}$\\
$\sigma_{3}/\sigma_{2}$ & 0.130${}^{+0.005}_{-0.007}$\\
$\sigma_{4}/\sigma_{3}$ & 0.13${}^{+0.02}_{-0.01}$\\
\end{tabular}
\end{ruledtabular}
\end{table}

\begin{table}[tbfh]
\caption{Inclusive hadronization correction derived with \textsc{sherpa} 1.2.3 and the CTEQ6.6 PDF set for each inclusive jet bin.}
\label{tb:XXX}
\begin{ruledtabular}
\begin{tabular}{l r}
Inclusive jet multiplicity bin & Hadronization correction and associated statistical uncertainty\\
\hline
 0-jet & 1.00\\
 1-jet & 1.021  $\pm$ 0.003\\
 2-jet & 1.052  $\pm$ 0.006\\
 3-jet & 1.035  $\pm$ 0.011\\
 4-jet & 1.078  $\pm$ 0.022\\
\end{tabular}
\end{ruledtabular}
\end{table}

\clearpage

\end{widetext}


\begin{thebibliography}{0}
\expandafter\ifx\csname natexlab\endcsname\relax\def\natexlab#1{#1}\fi
\expandafter\ifx\csname bibnamefont\endcsname\relax
  \def\bibnamefont#1{#1}\fi
\expandafter\ifx\csname bibfnamefont\endcsname\relax
  \def\bibfnamefont#1{#1}\fi
\expandafter\ifx\csname citenamefont\endcsname\relax
  \def\citenamefont#1{#1}\fi
\expandafter\ifx\csname url\endcsname\relax
  \def\url#1{\texttt{#1}}\fi
\expandafter\ifx\csname urlprefix\endcsname\relax\def\urlprefix{URL }\fi
\providecommand{\bibinfo}[2]{#2}
\providecommand{\eprint}[2][]{\url{#2}}

\end{thebibliography}


\begin{thebibliography}{99}

\bibitem{Aaltonen:2007ip}
  T.~Aaltonen {\it et al.}  [CDF Collaboration],
  %``Measurement of the cross section for $W^-$ boson production in association
  %with jets in ppbar collisions at $\sqrt{s}$ = 1.96-TeV,''
  Phys.\ Rev.\  D {\bf 77}, 011108 (2008).
  %[arXiv:0711.4044 [hep-ex]].

\bibitem{Berger:2009ep}
  C.~F.~Berger {\it et al.},
  %``Next-to-Leading Order QCD Predictions for W+3-Jet Distributions at Hadron
  %Colliders,''
  Phys.\ Rev.\  D {\bf 80}, 074036 (2009).
  %[arXiv:0907.1984 [hep-ph]].

\bibitem{Ellis:2009bu}
  R.~K.~Ellis, K.~Melnikov, and G.~Zanderighi,
  %``W+3 jet production at the Tevatron,''
  Phys.\ Rev.\  D {\bf 80}, 094002 (2009).
  %[arXiv:0906.1445 [hep-ph]].


%\bibitem{Mangano:2008ha}
%  M.~L.~Mangano,
%  %``Standard Model backgrounds to supersymmetry searches,''
%  Eur.\ Phys.\ J.\  C {\bf 59}, 373 (2009)
%  [arXiv:0809.1567 [hep-ph]].

\bibitem{Gleisberg:2008ta}
  T.~Gleisberg {\it et al.},
%  ``SHERPA v1.2.3",
  %``Event generation with SHERPA 1.1,''
  J.~High~Energy~Phys. {\bf 02}, 007 (2009).
  %[arXiv:0811.4622 [hep-ph]].

\bibitem{Abazov:2008ez}
  V.~M.~Abazov {\it et al.}  [D0 Collaboration],
  %``Measurement of differential $Z / \gamma^{*}$ + jet + $X$ cross sections in
  %$p \bar{p}$ collisions at $\sqrt{s}$ = 1.96-TeV,''
  Phys.\ Lett.\  B {\bf 669}, 278 (2008).
  %[arXiv:0808.1296 [hep-ex]].

\bibitem{Abazov:2010kn}
  V.~M.~Abazov {\it et al.}  [D0 Collaboration],
  %``Measurement of the normalized Z/gamma*->mu+mu- transverse momentum
  %distribution in p\bar{p} collisions at sqrt{s}=1.96 TeV,''
  Phys.\ Lett.\  B {\bf 693}, 522 (2010).
  %[arXiv:1006.0618 [hep-ex]].

\bibitem{particle}
  C.~Buttar {\it et al.}, 
  arXiv:0803.0678 [hep-ph], Section 9.

\bibitem{d0det}
  V.~M.~Abazov {\it et al.} [D0 Collaboration],
  Nucl.\ Instrum.\ Methods Phys.\ Res.\ A {\bf 565}, 463  (2006).

 \bibitem{d0jets}
  G.~C.~Blazey {\it et al.}, 
  % in {\sl Proceedings of the Workshop: QCD and Weak Boson Physics in Run II,} edited by U.~Baur, R.K.~Ellis, and D. Zeppenfeld, 
  Fermilab-Pub-00/297 (2000).

\bibitem{definitions}
We use a standard right-handed coordinate system. The nominal collision point is the center of the detector with coordinate $(0,0,0)$. The direction of the proton beam is the positive $+z$ axis. The $+x$ axis is horizontal, pointing away from the center of the Tevatron ring. The $+y$ axis points vertically upwards. The polar angle, $\theta$, is defined such that $\theta = 0$ is the $+z$ direction.  The rapidity is defined as $y$ = $-\ln[(E + p_Z )/(E - p_Z )]$, where $E$ is the energy and $p_Z$ is the momentum component along the proton beam direction.
Pseudorapidity is defined as $\eta$ = $-\ln[\tan\frac{\theta}{2}]$. $\phi$ is defined as the azimuthal angle in the plane transverse to the proton beam direction.

\bibitem{JES}
  V.~M.~Abazov {\it et al.}  [D0 Collaboration],
  %  ``Measurement of the inclusive jet cross-section in $p \bar{p}$ collisions at
  %$s^{91/2)}$ =1.96-TeV,''
  Phys.\ Rev.\ Lett.\  {\bf 101}, 062001 (2008)
  %[arXiv:0802.2400 [hep-ex]].
  %%CITATION = PRLTA,101,062001;%%

\bibitem{Mangano:2002ea}
  M.~L.~Mangano {\it et al.},
  %``ALPGEN, a generator for hard multiparton processes in hadronic
  %collisions,''
  J.~High~Energy~Phys. {\bf 07}, 001 (2003).
  We use version 2.11.
  %[arXiv:hep-ph/0206293].

\bibitem{pythia}
    %{\sc Pythia} reference: \\
    T. Sj\"{o}strand {\it et al.}, Comput. Phys. Commun. {\bf 135}, 238 (2001).
    We use version 6.403.

\bibitem{Boos:2004kh}
  E.~Boos {\it et al.}  [CompHEP Collaboration],
  %``CompHEP 4.4: Automatic computations from Lagrangians to events,''
  Nucl.\ Instrum.\ Meth.\ Phys\ Res.~A {\bf 534}, 250 (2004).
  %[arXiv:hep-ph/0403113].

\bibitem{MCFM}
  J. Campbell and R.K. Ellis, 
  Phys.\ Rev.\ D {\bf 65}, 113007 (2002);
  J. Campbell, R.K. Ellis, and D. Rainwater, 
  Phys.\ Rev.\ D {\bf 68}, 094021 (2003).

\bibitem{Abazov:2007kg}
  V.~M.~Abazov {\it et al.}  [D0 Collaboration],
  %``Measurement of the $t \bar{t}$ production cross section in $p \bar{p}$
  %collisions at $\sqrt{s}$ = 1.96-TeV using kinematic characteristics of lepton
  %+ jets events,''
  Phys.\ Rev.\  D {\bf 76}, 092007 (2007).
  %[arXiv:0705.2788 [hep-ex]].

\bibitem{Hocker:1995kb}
  A.~Hocker and V.~Kartvelishvili,
  %``SVD approach to data unfolding,''
  Nucl.\ Instrum.\ Meth.\ Phys.\ Res.\ A~{\bf 372} (1996).
  %[hep-ph/9509307].

\bibitem{Berger:2009zg}
  C.~F.~Berger {\it et al.},
  %``Precise Predictions for $W$ + 3 Jet Production at Hadron Colliders,''
  Phys.\ Rev.\ Lett.\  {\bf 102}, 222001 (2009).
  %[arXiv:0902.2760 [hep-ph]].

\bibitem{Berger:2010zx}
  C.~F.~Berger {\it et al.},
  %``Precise Predictions for W + 4 Jet Production at the Large Hadron
  Phys.\ Rev.\ Lett.\  {\bf 106}, 092001 (2011).
  %Collider,''
  %arXiv:1009.2338 [hep-ph].

\bibitem{Berger:2009ba}
  C.~F.~Berger {\it et al.},
  %``Multi-jet cross sections at NLO with BlackHat and Sherpa,''
  arXiv:0905.2735 [hep-ph].
  %%CITATION = ARXIV:0905.2735;%%

\bibitem{Ellis:2008qc}
  R.~K.~Ellis {\it et al.},
  %``One-loop amplitudes for $W^+$ 3 jet production in hadron collisions,''
  J.~High~Energy~Phys. {\bf 01}, 012 (2009).
  %[arXiv:0810.2762 [hep-ph]].
  %%CITATION = JHEPA,0901,012;%%

\bibitem{Giele:2008bc}
  W.~T.~Giele and G.~Zanderighi,
  %``On the Numerical Evaluation of One-Loop Amplitudes: The Gluonic Case,''
  J.~High~Energy~Phys. {\bf 06}, 038 (2008).
  %[arXiv:0805.2152 [hep-ph]].
  %%CITATION = JHEPA,0806,038;%%

\bibitem{Salam:2007xv}
  G.~P.~Salam and G.~Soyez,
%  %``A practical Seedless Infrared-Safe Cone jet algorithm,''
  J.~High~Energy~Phys. {\bf 05}, 086 (2007).
  %[arXiv:0704.0292 [hep-ph]].

\bibitem{Bauer:2009km}
  C.~W.~Bauer and B.~O.~Lange,
  %``Scale setting and resummation of logarithms in pp ---> V + jets,''
  arXiv:0905.4739 [hep-ph].

\bibitem{mstw}
    A. D. Martin {\it et al.},
    Eur. Phys. J. C {\bf 63}, 189 (2009).

\bibitem{cteq}
%    {\sc Cteq6} reference: \\
    J. Pumplin {\it et al.}, J.~High~Energy~Phys. {\bf 07}, 012 (2002);
    D. Stump {\it et al.}, J.~High~Energy~Phys. {\bf 10}, 046 (2003).

\bibitem{WTSYDP}
  G.~D.~Lafferty and T.~R.~Wyatt,
  %``Where to stick your data points: The treatment of measurements within wide
  %bins,''
  Nucl.\ Instrum.\ Meth.\  A {\bf 355} (1995) 541.
  %%CITATION = NUIMA,A355,541;%%


\bibitem{Abazov:2009av}
  V.~M.~Abazov {\it et al.}  [D0 Collaboration],
  %``Measurements of differential cross sections of $Z /\gamma^\ast$+jets+X
  %events in proton anti-proton collisions at $\sqrt{s}$=1.96 TeV,''
  Phys.\ Lett.\  B {\bf 678}, 45 (2009).
  %[arXiv:0903.1748 [hep-ex]].

\bibitem{appendix}
See the Appendix for supplementary material including tabulated $W$+$n$ jet cross section measurements, theoretical 
predictions, and hadronization corrections applied to the theoretical predictions.

\bibitem{Abazov:2009pp}
  V.~M.~Abazov {\it et al.}  [D0 Collaboration],
  %``Measurement of $Z / \gamma^\ast +jet+X$ angular distributions in $p
  %\bar{p}$ collisions at $\sqrt{s}=1.96$ TeV,''
  Phys.\ Lett.\  B {\bf 682}, 370 (2010).
  %[arXiv:0907.4286 [hep-ex]].

 

\end{thebibliography}
\end{document}
%
% ****** End of file template.aps ******